\documentclass[amstex,twocolumn,showpacs,floats,floatfix,superscriptaddress,aps,pra]{revtex4}
\usepackage{amssymb}
\usepackage{amsmath}
\usepackage{calc}
\usepackage{graphicx}
\usepackage{bm}

\begin{document}

\author{G. S. Vasilev}
\affiliation{Department of Physics, Sofia University, James Bourchier 5 blvd, 1164 Sofia, Bulgaria}
\author{S. S. Ivanov}
\affiliation{Department of Physics, Sofia University, James Bourchier 5 blvd, 1164 Sofia, Bulgaria}
\author{N. V. Vitanov}
\affiliation{Department of Physics, Sofia University, James Bourchier 5 blvd, 1164 Sofia, Bulgaria}
\affiliation{Institute of Solid State Physics, Bulgarian Academy of Sciences, Tsarigradsko chauss\'{e}e 72, 1784 Sofia, Bulgaria}
\title{Degenerate Landau-Zener model: Exact analytical solution}
\date{\today }

\begin{abstract}
The exact analytical solution of the degenerate Landau-Zener model, wherein two bands of degenerate energies cross in time, is presented. 
The solution is derived by using the Morris-Shore transformation, which reduces the fully coupled system
 to a set of independent nondegenerate two-state systems and a set of decoupled states. 
Due to the divergence of the phase of the off-diagonal element of the propagator in the original Landau-Zener model,
 not all transition probabilites exist for infinite time duration. 
In general, apart from some special cases, only the transition probabilities between states within the same degenerate set exist,
 but not between states of different sets.
An illustration is presented for the transition between the magnetic sublevels of two atomic levels with total angular momenta $J=2$ and 1. 
\end{abstract}
\pacs{32.80.Bx, 33.80.Be, 03.65.Ge, 34.70.+e}
\maketitle


\section{Introduction}


The Landau-Zener (LZ) model \cite{LZ} is the most popular tool for estimating the transition probability between two states whose energies
 cross in time, a situation which can occur in virtually every area of quantum physics. 
The LZ Hamiltonian involves the simplest nontrivial time dependence: a constant interaction and linearly changing energies.
Nevertheless, due to some mathematical subtleties, when applied to real physical systems with more sophisticated time dependences the LZ model
 often provides more accurate results than anticipated.

The popularity of the LZ model, largely ensuing from the extreme simplicity of the transition probability,
 has stimulated numerous extensions to multiple levels. 
There are two main types of generalizations: single-crossing bow-tie models and multiple-crossings grid models.

In the bow-tie models all energies cross at the same instant of time.
Analytic bow-tie solutions have been found for three \cite{Carroll} and $N$ states \cite{Ostrovsky97,Harmin,Brundobler}. 
Examples of such systems occur, for instance, in a rf-pulse controlled Bose-Einstein condensate output coupler \cite{Mewes,VitanovBEC},
 and in the coupling pattern of Rydberg sublevels in a magnetic field \cite{Harmin}.
An extension, where one of the levels is split into two parallel levels, has been solved by Demkov and Ostrovsky \cite{Demkov01}. 

In the multiple-crossings models, a set of $N_{a}$ parallel equidistant linear energies cross another set of $N_{b}$ such energies,
 thus forming a grid of crossings (Demkov-Ostrovsky model) \cite{Demkov95,Usuki,Ostrovsky98}. 
For $N_{b}=1$ (or $N_{a}=1$) the Demkov-Ostrovsky model reduces to the earlier Demkov-Osherov model \cite{Demkov68,Kayanuma}. 
The special case when $N_{a}$ and $N_{b}$ are infinite (so that the grid of crossings is fully periodic) has also been solved \cite{Demkov95b}. 
In the most general case of an arbitrary linear Hamiltonian, $\mathbf{H}(t)=\mathbf{D}+\mathbf{C}t$, where $\mathbf{C}$ is diagonal,
 the general solution has not been derived yet, but exact results for some survival probabilities have been conjectured \cite{Brundobler}
 and derived \cite{Shytov,Sinitsyn04,Volkov04,Volkov05}.

A variety of physical systems provide examples of multiple level crossings.
Among them we mention ladder climbing of atomic and molecular states by chirped laser pulses \cite{Melinger,ARPC},
 harpoon model for reactive scattering \cite{Child}, and optical shielding in cold atomic collisions \cite{shielding}.

A general feature of all soluble multilevel crossing models is that the transition probabilities $P_{m\rightarrow n}$ between states
 $\psi _{m}$ and $\psi _{n}$ are given by very simple expressions, as in the original LZ model, although the derivations are not trivial. 
In the grid models, in particular, the exact probabilities $P_{m\rightarrow n}$ have the same form (products of LZ probabilities
 for transition or no-transition applied at the relevant crossings) as what would be obtained by naive multiplication of LZ probabilities
 while moving across the grid of crossings from $\psi _{m}$ to $\psi _{n}$, without accounting for phases and interferences. 
For instance, the counterintuitive transitions, for which the level crossings appear in a ``wrong'' order in time, are forbidden at infinite times. 
It has been shown, though, that the probability for counterintuitive transitions is nonzero for finite interaction duration
 \cite{Rangelov} or for piecewise-linear sloped potential \cite{Yurovsky99}.

An interesting feature of the existing multistate LZ solutions is that the respective derivations (usually using Laplace transforms
 and contour integration) all fail in the limit of \emph{degenerate levels} and the assumption of nondegeneracy is essential. 
Effects of level degeneracies in the Demkov-Osherov model have been studied by reducing the multistate dynamics
 to that of a single nondegenerate two-state system and several decoupled states \cite{Yurovsky98,Kyoseva}. 
Effects of quasi-degeneracies have been described by treating a nondegenerate system with small energy gaps as a perturbed degenerate system \cite{Yurovsky99}.

In this paper, we derive the exact analytical solution for two crossing degenerate levels $a$ and $b$,
 of arbitrary degeneracies $N_{a}$ and $N_{b}$, which we shall refer to as the \emph{degenerate LZ model}. 
Our model can therefore be considered as an extension of the standard nondegenerate two-state LZ model to two degenerate levels. 
It also generalizes the solutions by Yurovsky and Ben-Reuven \cite{Yurovsky98} and by Kyoseva and Vitanov \cite{Kyoseva},
 which assume one degenerate and one nondegenerate level. 
This model can also be viewed as the unsolved limiting case of the Demkov-Ostrovsky model \cite{Demkov95,Usuki,Ostrovsky98} for vanishing level spacing. 
Finally, this model represents the unsolved limiting case of the bow-tie models \cite{Ostrovsky97,Harmin,Brundobler,Demkov01}
 when all energies, that cross at the same time, coalesce into only two different slopes.

Our method of solution is drastically different, and much simpler than those used in the nondegenerate multistate LZ models. 
We make use of the powerful Morris-Shore (MS) transformation, which reduces the dynamics of two sets of
 degenerate states into that of a collection of independent nondegenerate two-state systems and decoupled (dark) states. 
Each of the independent two-state systems represents a standard, nondegenerate LZ problem, whereas the decoupled states do not evolve. 
Hence the solution of the degenerate LZ problem is equivalent to a collection of two-state LZ solutions. 
However, the situation is not so trivial because the different LZ solutions interfere
 and produce interesting features in the probabilities in the original basis.
In particular, it turns out that not all transition probabilities are defined, as far as an infinite interaction duration is concerned.

Among the numerous possible physical realizations of the degenerate LZ model, we point out the degenerate two-level system formed between two
 atomic levels of angular momenta $J_{a}$ and $J_{b}=J_{a}$ or $J_{a}\pm 1$, driven by linearly chirped laser fields of arbitrary polarizations. 
In the absence of magnetic field such a system represents exactly a degenerate LZ model.

This paper is organized as follows. We define the problem in Sec. \ref{Sec-definition} and the propagator is derived in Sec. \ref{Sec-solution}
 in the general case. 
A special example for a $J_{a}=2\leftrightarrow J_{b}=1$ transition is considered in Sec. \ref{Sec-example}. 
The conclusions are summarized in Sec. \ref{Sec-conclusions}.


\section{Definition of the degenerate Landau-Zener model\label{Sec-definition}}


\begin{figure}[tb]
\includegraphics[width=72mm]{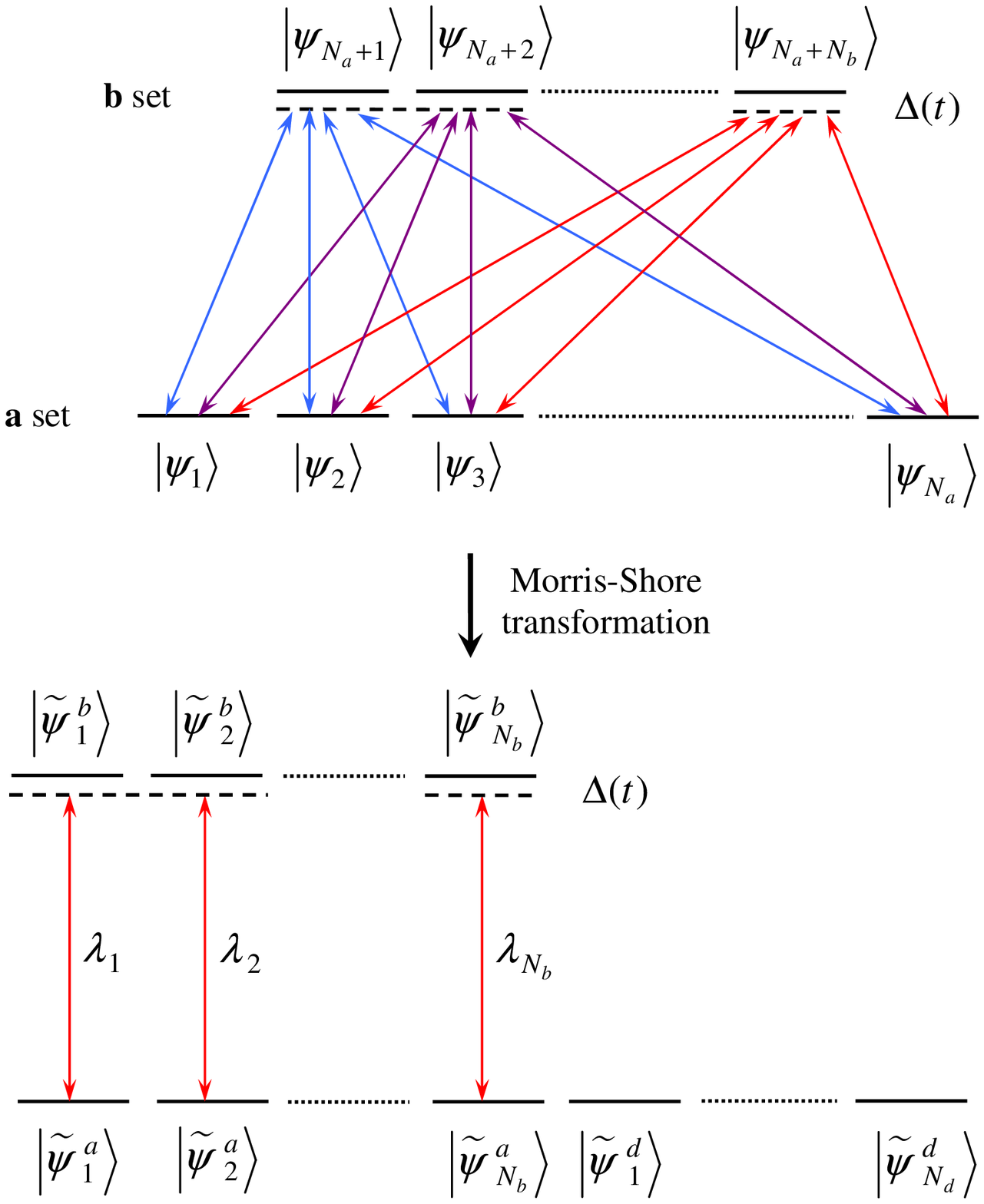}
\caption{(Color online) The Morris-Shore transformation: a multistate system consisting of two coupled sets of degenerate levels is
decomposed into a set of independent nondegenerate two-state systems and a set of decoupled states.}
\label{Fig-system}
\end{figure}

We consider a quantum system with $N_a$ degenerate states $\left\{\vert \psi_{m}\rangle \right\} _{m=1}^{N_{a}}$ in the (lower) $a$ set
 and $N_b$ states $\left\{\vert\psi_{N_{a}+n}\rangle \right\}_{n=1}^{N_b}$ in the (upper) $b$ set, as displayed in Fig. \ref{Fig-system} (top). 
Without loss of generality we assume that $N_{a}\geqq N_{b}$. 
Each of the $a$ states $\left\vert \psi _{m}\right\rangle $ is coupled to each of the $b$ states $\left\vert \psi _{n}\right\rangle $
 by a constant coupling $\Omega _{mn}$, and all couplings can be different. 
The $a$ states are not coupled to each other directly, neither are the $b$ states.
All fields are off resonance by the same detuning $\Delta (t)$, which is assumed to be linear in time,
 with a rate $C$ (\emph{chirp} in coherent atomic excitation \cite{Shore,AE}), 
\begin{subequations}
\begin{eqnarray}
\Omega _{mn} &=&\text{const},  \label{couplings} \\
\Delta (t) &=&Ct,  \label{detunings}
\end{eqnarray}
\end{subequations}

For $N_{b}=1$ the present model reduces to the $N$-pod model solved by Kyoseva and Vitanov \cite{Kyoseva},
 and for $N_{a}=N_{b}=1$ to the nondegenerate original LZ model. 
The present model is therefore a generalization of the $N$-pod model to a degenerate $b$ level.

We adopt a state ordering wherein the $N_{a}$ sublevels of the $a$ level are placed first, followed by the $N_{b}$ sublevels of the $b$ level. 
In the rotating-wave approximation (RWA) the Schr\"{o}dinger equation of the system reads \cite{Shore} 
\begin{equation}
i\hbar \frac{d}{dt}\mathbf{C}(t)=\mathbf{H(}t)\mathbf{C}(t),  \label{Sch-eq}
\end{equation}%
where the elements of the ($N_{a}+N_{b}$)-dimensional vector $\mathbf{C}(t)$
are the probability amplitudes of the states. The adopted state ordering
allows us to write the RWA Hamiltonian as a block matrix,%
\begin{equation}
\mathbf{H(}t)=\left[ 
\begin{array}{cc}
\mathbf{0} & \mathbf{V} \\ 
\mathbf{V}^{\dagger } & \mathbf{D}(t)%
\end{array}%
\right] ,  \label{H}
\end{equation}%
Here $\mathbf{0}$ is the $N_{a}$-dimensional square zero matrix, in which
the zero off-diagonal elements indicate the absence of couplings between the 
$a$ states, while the zero diagonal elements show that the $a$ states have
the same energy, which is taken as the zero of the energy scale. The matrix $%
\mathbf{D}(t)$ is an $N_{b}$-dimensional square diagonal matrix, with $%
\Delta (t)$ on the diagonal, $\mathbf{D}(t)=\Delta (t)\mathbf{1}_{N_{b}}$.
The absence of off-diagonal elements in $\mathbf{D}$ reflects again the
absence of couplings between the $b$ states, while the diagonal elements $%
\Delta $ stand for the common energy of all $b$ states.

In Eq. (\ref{H}) $\mathbf{V}$ is an $\left( N_{a}\times N_{b}\right) $-dimensional interaction matrix with constant elements, 
\begin{equation}
\mathbf{V} \!=\! \left[\! \begin{array}{cccc}
\Omega _{11} & \Omega _{12} & ... & \Omega _{1N_{b}} \\ 
\Omega _{21} & \Omega _{22} & ... & \Omega _{2N_{b}} \\ 
\vdots & \vdots & ... & \vdots \\ 
\Omega _{N_{a}1} & \Omega _{N_{a}2} & ... & \Omega _{N_{a}N_{b}}%
\end{array}\! \right] \!
=\left[ \vert \Omega _{1}\rangle ,\vert \Omega_{2}\rangle ,\ldots ,\vert \Omega _{N_{b}}\rangle \right] ,
\label{V}
\end{equation}%
where $\left\vert \Omega _{n}\right\rangle $ ($n=1,2,\ldots ,N_{b}$) are $N_{a}$-dimensional vectors
 comprising the interactions of the $n$th state of the $b$ set with all states of the $a$ set, 
\begin{equation}
\left\vert \Omega _{n}\right\rangle =\left[ 
\begin{array}{c}
\Omega _{1n} \\ 
\Omega _{2n} \\ 
\vdots \\ 
\Omega _{N_{a}n}%
\end{array}%
\right] \quad (n=1,2,\ldots ,N_{b}).  \label{Omega_n}
\end{equation}


\section{Exact analytic solution of the degenerate Landau-Zener model\label{Sec-solution}}


\subsection{Morris-Shore transformation}


We shall solve the degenerate LZ problem by using the Morris-Shore (MS) transformation \cite{Morris83}. 
Morris and Shore have shown that any degenerate two-level system, in which all couplings share the same time dependence (constant in our case)
 and the same detuning (linear here), can be reduced with a constant unitary transformation $\mathbf{S}$ to an equivalent system
 comprising only independent two-state systems and uncoupled (dark) states, as shown in Fig. \ref{Fig-system}. 
This transformation reads 
\begin{equation}
|\psi _{i}\rangle =\sum_{k}S_{ki}|\widetilde{\psi }_{k}\rangle \quad
\Longleftrightarrow \quad |\widetilde{\psi }_{k}\rangle =\sum_{i}S_{ki}^{\ast }|\psi_{i}\rangle ,  \label{transformation}
\end{equation}%
where the tildas denote the MS basis hereafter. 
The constant transformation matrix $\mathbf{S}$ can be represented in the block-matrix form 
\begin{equation}
\mathbf{S}=\left[ 
\begin{array}{cc}
\mathbf{A} & \mathbf{O} \\ 
\mathbf{O} & \mathbf{B}%
\end{array}%
\right] ,  \label{W}
\end{equation}%
where $\mathbf{A}$ is a unitary $N_{a}$-dimensional square matrix and $\mathbf{B}$ is a unitary $N_{b}$-dimensional square matrix,
 $\mathbf{AA}^\dagger=\mathbf{A}^\dagger\mathbf{A}=\mathbf{1}_{N_a}$ and $\mathbf{BB}^\dagger=\mathbf{B}^\dagger\mathbf{B}=\mathbf{1}_{N_b}$. 
The constant matrices $\mathbf{A}$ and $\mathbf{B}$ mix only sublevels of a given level:
 $\mathbf{A}$ mixes the $a$ sublevels and $\mathbf{B}$ mixes the $b$ sublevels. 
The transformed MS Hamiltonian has the form%
\begin{equation}
\widetilde{\mathbf{H}}(t)=\mathbf{SH}(t)\mathbf{S}^{\dagger }=\left[ 
\begin{array}{cc}
\mathbf{O} & \widetilde{\mathbf{V}} \\ 
\widetilde{\mathbf{V}}^{\dagger } & \mathbf{D}(t)%
\end{array}%
\right] ,  \label{H-MS}
\end{equation}%
where 
\begin{equation}
\widetilde{\mathbf{V}}=\mathbf{AVB}^{\dagger }.  \label{M}
\end{equation}%
The $N_a\times N_b$ matrix $\widetilde{\mathbf{V}}$ has $N_d=N_a-N_b$ null rows ($N_a\geqq N_b$), which correspond to decoupled states. 
The decomposition of $\mathbf{H}$ into a set of independent two-state systems requires that, after removing the null rows,
 $\widetilde{\mathbf{V}}$ reduces to a $N_b$-dimensional diagonal matrix; let us denote its diagonal elements by $\lambda_n$ ($n=1,2,\ldots,N_b$).

It follows from Eq. (\ref{M}) that 
\begin{subequations}
\label{MM}
\begin{eqnarray}
\widetilde{\mathbf{V}}\widetilde{\mathbf{V}}^{\dagger } &=&\mathbf{AVV}^{\dagger }\mathbf{A}^{\dagger },  \label{MM+} \\
\widetilde{\mathbf{V}}^{\dagger }\widetilde{\mathbf{V}} &=&\mathbf{BV}^{\dagger }\mathbf{VB}^{\dagger }.  \label{M+M}
\end{eqnarray}%
Hence $\mathbf{A}$ and $\mathbf{B}$ are defined by the condition that they diagonalize $\mathbf{VV}^{\dagger }$
 and $\mathbf{V}^{\dagger }\mathbf{V}$, respectively. 
It is important to note that the square matrices $\mathbf{VV}^{\dagger }$ and $\mathbf{V}^{\dagger }\mathbf{V}$ have different dimensions,
 $N_{a}$ and $N_{b}$, respectively. 
Because all elements of $\mathbf{V}$ are constant, $\mathbf{A}$ and $\mathbf{B}$ are also constant.
It is straightforward to show that the $N_{b}$ eigenvalues of $\mathbf{V}^{\dagger }\mathbf{V}$ are all non-negative;
 according to Eqs. (\ref{M}) and (\ref{MM}) they are $\lambda _{n} ^{2}$ ($n=1,2,\ldots ,N_{b}$). 
The matrix $\mathbf{VV}^{\dagger }$ has the same eigenvalues and additional $N_{d}=N_{a}-N_{b}$ zero eigenvalues.

The MS Hamiltonian (\ref{H-MS}) has the explicit form 
\end{subequations}
\begin{equation}
\widetilde{\mathbf{H}}=\left[ \text{ }%
\begin{tabular}{cc}
$\mathbf{0}_{N_{d}}$ & $\mathbf{0}$ \\ 
$\mathbf{0}$ & $\!\!\!%
\begin{array}{cccccccc}
0 & 0 & \cdots & 0 & \lambda _{1} & 0 & \cdots & 0 \\ 
0 & 0 & \cdots & 0 & 0 & \lambda _{2} & \cdots & 0 \\ 
\vdots & \vdots & \ddots & \vdots & \vdots & \vdots & \ddots & \vdots \\ 
0 & 0 & \cdots & 0 & 0 & 0 & \cdots & \lambda _{N_{b}} \\ 
\lambda _{1} & 0 & \cdots & 0 & \Delta & 0 & \cdots & 0 \\ 
0 & \lambda _{2} & \cdots & 0 & 0 & \Delta & \cdots & 0 \\ 
\vdots & \vdots & \ddots & \vdots & \vdots & \vdots & \ddots & \vdots \\ 
0 & 0 & \cdots & \lambda _{N_{b}} & 0 & 0 & \cdots & \Delta%
\end{array}%
$%
\end{tabular}%
\!\text{ }\right] .  \label{H-MS explicit}
\end{equation}%
The structure of $\widetilde{\mathbf{H}}$ shows that in the MS basis the dynamics is decomposed into sets of $N_{d}$ decoupled single states,
 and $N_{b}$ independent two-state systems $|\widetilde{\psi}_{n}^{a}\rangle \leftrightarrow |\widetilde{\psi }_{n}^{b}\rangle$
 ($n=1,2,\ldots ,N_b$), each composed of an $a$ state $|\widetilde{\psi}_n^a\rangle$ and a $b$ state $|\widetilde{\psi}_n^b\rangle$,
 and driven by the Hamiltonians, 
\begin{equation}
\widetilde{\mathbf{H}}_{n}(t)=\left[ \begin{array}{cc}
0 & \lambda _{n} \\ 
\lambda _{n} & \Delta (t)%
\end{array}\right] \quad (n=1,2,\ldots ,N_{b}).  \label{Hn}
\end{equation}%
These two-state Hamiltonians have the same detuning $\Delta (t)$ but different couplings $\lambda _{n}$. 
Each of the new $a$ states $|\widetilde{\psi }_{n}^{a}\rangle $ is the eigenstate of $\mathbf{VV}^{\dagger }$
 corresponding to the eigenvalue $\lambda _{n}^{2}$, whereas each of the new $b$ states $|\widetilde{\psi }_{n}^{b}\rangle $ is the
 eigenstate of $\mathbf{V}^{\dagger }\mathbf{V}$, corresponding to the same eigenvalue $\lambda _{n}^{2}$. 
The square root of this common eigenvalue, $\lambda _{n}$, represents the coupling between
 $|\widetilde{\psi }_{n}^{a}\rangle$ and $|\widetilde{\psi }_{n}^{b}\rangle $. 
The $N_{d}$ zero eigenvalues of $\mathbf{VV}^{\dagger }$ correspond to decoupled (dark) states in the $a$ set
 (since we assume that $N_{a}\geqq N_{b}$, dark states, if any, are in the $a$ set). 
The dark states are decoupled from the dynamical evolution because they are driven by one-dimensional null Hamiltonians.


\subsection{Solution to the degenerate LZ problem}


\subsubsection{The MS transformation}

The MS decomposition allows us to reduce the degenerate two-level LZ problem to a set of nondegenerate two-state LZ problems,
 wherein the detuning is unchanged and given by Eq. (\ref{detunings})
 while the couplings $\lambda _{n}$, defined as the square roots of the eigenvalues of $\mathbf{V}^{\dagger }\mathbf{V}$,
 are combinations of the initial couplings between the $a$ and $b$ states.

From the vector form (\ref{V}) of $\mathbf{V}$ we obtain 
\begin{subequations}
\begin{eqnarray}
\mathbf{VV}^{\dagger } &=&\sum_{n=1}^{N_{b}}\left\vert \Omega
_{n}\right\rangle \left\langle \Omega _{n}\right\vert ,  \label{VV+} \\
\mathbf{V}^{\dagger }\mathbf{V} &=&\left[ 
\begin{array}{cccc}
\left\langle \Omega _{1}|\Omega _{1}\right\rangle & \left\langle \Omega
_{1}|\Omega _{2}\right\rangle & \cdots & \left\langle \Omega _{1}|\Omega
_{N_{b}}\right\rangle \\ 
\left\langle \Omega _{2}|\Omega _{1}\right\rangle & \left\langle \Omega
_{2}|\Omega _{2}\right\rangle & \cdots & \left\langle \Omega _{2}|\Omega
_{N_{b}}\right\rangle \\ 
\vdots & \vdots & \ddots & \vdots \\ 
\left\langle \Omega _{N_{b}}|\Omega _{1}\right\rangle & \left\langle \Omega
_{N_{b}}|\Omega _{2}\right\rangle & \cdots & \left\langle \Omega
_{N_{b}}|\Omega _{N_{b}}\right\rangle%
\end{array}%
\right] .  \label{V+V}
\end{eqnarray}%
Note that $\mathbf{V}^\dagger\mathbf{V}$ is the Gram matrix for the set of vectors $\left\{\vert\Omega_n\rangle\right\}_{n=1}^{N_{b}}$.
Thus if all these vectors are linearly independent then $\det \mathbf{V}^{\dagger }\mathbf{V}\neq 0$ and all eigenvalues of
 $\mathbf{V}^{\dagger }\mathbf{V}$ are nonzero \cite{Gantmacher}; however, this assumption is unnecessary.

We assume that we can find the eigenvalues $\lambda _{n}^{2}$ $(n=1,2,\ldots ,N_{b})$ of the matrices (\ref{VV+}) and (\ref{V+V}),
 and the corresponding orthonormalized eigenvectors: the $N_b$ coupled eigenstates $|\widetilde{\psi }_{n}^{a}\rangle $
 of $\mathbf{VV}^{\dagger }$ and $|\widetilde{\psi }_{n}^{b}\rangle $ of $\mathbf{V}^{\dagger }\mathbf{V}$,
 and the $N_d$ decoupled eigenstates $|\widetilde{\psi }_{k}^{d}\rangle $ of $\mathbf{VV}^{\dagger }$. 
We use these eigenstates to construct the transformation matrices as  
\end{subequations}
\begin{equation}
\mathbf{A}=\left[ 
\begin{array}{c}
\langle \widetilde{\psi }_{1}^{d}| \\ 
\vdots \\ 
\langle \widetilde{\psi }_{N_{d}}^{d}| \\ 
\langle \widetilde{\psi }_{1}^{a}| \\ 
\vdots \\ 
\langle \widetilde{\psi }_{N_{b}}^{a}|%
\end{array}%
\right] ,\quad \mathbf{B}=\left[ 
\begin{array}{c}
\langle \widetilde{\psi }_{1}^{b}| \\ 
\vdots \\ 
\langle \widetilde{\psi }_{N_{b}}^{b}|%
\end{array}%
\right] .  \label{A B}
\end{equation}%
Then according to the general theory the transformed interaction matrix (\ref{M}) in the MS basis takes the form (\ref{H-MS explicit}),
 where the positions of the $N_{d}$ zero eigenvalues and the $N_{b}$ eigenvalues $\lambda _{n}$
 are determined by the ordering of the eigenstates in the transformation matrices (\ref{A B}).

\subsubsection{The MS propagators}

Because the dark states are decoupled and have zero energies, their propagator is the unit matrix $\mathbf{1}_{N_{d}}$.

The propagator for each of the two-state MS Hamiltonians (\ref{Hn}) is the LZ propagator for the respective coupling $\lambda _{n}$,%
\begin{subequations}\label{Un}
\begin{eqnarray}
\widetilde{\mathbf{U}}_{n} &=&e^{-i\delta /2}\left[ 
\begin{array}{cc}
\alpha _{n} & -\beta _{n}^{\ast } \\ 
\beta _{n} & \alpha _{n}^{\ast }%
\end{array}%
\right] ,  \label{Un1} \\
\delta &=&\int_{t_{i}}^{t_{f}}\Delta(t) dt=\frac{1}{2}\left( \tau _{f}^{2}-\tau_{i}^{2}\right) .\label{delta}
\end{eqnarray}
\end{subequations}
The Cayley-Klein parameters are \cite{Finite LZ} 
\begin{subequations}
\label{finite LZ}
\begin{eqnarray}
\alpha &=&\frac{\Gamma (1-i\kappa ^{2})}{\sqrt{2\pi }}\left[ D_{i\kappa
^{2}}(\tau _{f}e^{-i\pi /4})D_{i\kappa ^{2}-1}(\tau _{i}e^{3i\pi /4})\right.
\notag \\
&&\left. +D_{i\kappa ^{2}}(\tau _{f}e^{3i\pi /4})D_{i\kappa ^{2}-1}(\tau
_{i}e^{-i\pi /4})\right] ,  \label{alpha finite} \\
\beta &=&\frac{\Gamma (1-i\kappa ^{2})}{\kappa \sqrt{2\pi }}e^{i\pi /4}\left[
-D_{i\kappa ^{2}}(\tau _{f}e^{-i\pi /4})D_{i\kappa ^{2}}(\tau _{i}e^{3i\pi
/4})\right.  \notag \\
&&\left. +D_{i\kappa ^{2}}(\tau _{f}e^{3i\pi /4})D_{i\kappa ^{2}}(\tau
_{i}e^{-i\pi /4})\right] ,  \label{beta finite}
\end{eqnarray}%
\end{subequations}
where $\kappa =\Omega _{0}/\sqrt{C}$, $\tau =t\sqrt{C}$, and $D_{v}(z)$ is the parabolic-cylinder function.
$\tau _{i}=t_i\sqrt{C}$ and $\tau _{f}=t_f\sqrt{C}$ are the scaled initial and final times, respectively. 
In the original LZ model, $\tau _{i}\rightarrow -\infty $ and $\tau _{f}\rightarrow \infty $,
 and the Cayley-Klein parameters read \cite{Finite LZ} 
\begin{subequations}
\begin{eqnarray}
\alpha _{n} &=&e^{-\pi \Lambda _{n}},  \label{alpha_n} \\
\beta _{n} &=&-e^{i\phi _{n}}\sqrt{1-e^{-2\pi \Lambda _{n}}},  \label{beta_n}
\end{eqnarray}%
\end{subequations}
with 
\begin{subequations}\label{LZ parameters}
\begin{eqnarray}
\Lambda _{n} &=&\frac{\lambda _{n}^{2}}{C},  \label{Ln} \\
\phi _{n} &=&\frac{\tau _{i}^{2}+\tau _{f}^{2}}{4}+\frac{1}{2}\Lambda
_{n}\ln \left( \tau _{i}^{2}\tau _{f}^{2}\right) +\phi _{n}^{LZ},
\label{phi_n} \\
\phi _{n}^{LZ} &=&\frac{\pi }{4}+\arg \Gamma \left( 1-i\Lambda _{n}\right) .
\label{phi_n^lz}
\end{eqnarray}
\end{subequations}
Hence the phase $\phi _{n}$ diverges, which is a result of the unphysical assumption
 of an infinitely long interaction duration. 
This divergence is unimportant in the original LZ model because the transition probability, 
\begin{equation}
P_{n}=\left\vert \beta _{n}\right\vert ^{2}=1-e^{-2\pi \Lambda _{n}},
\label{P-LZ}
\end{equation}%
is well defined.
Hence the final populations are well defined if the system starts in one of the two states, which is usually the case. 
However, when the system starts in a \emph{superposition} of states,
 this divergence does not allow to calculate the populations, even in the original LZ model.
We shall show below that in the degenerate LZ model this divergence does not allow for
 definite values of some populations even when the system starts in a single state.

There are two divergent terms in the phase (\ref{phi_n}): polynomial and logarithmic, with different origins and different implications. 
The term $\frac{1}{4}(\tau _{i}^{2}+\tau _{f}^{2})$ is unimportant in the present context because it derives from the chosen
 Schr\"{o}dinger representation (\ref{H}); in the interaction representation (when the detunings turn into phase factors of the couplings) it disappears. 
Moreover, this term is the same for all $\beta _{n}$ and factors out of the probabilities (see below). 
The term $\frac{1}{2}\Lambda _{n}\ln (\tau _{i}^{2}\tau _{f}^{2})$, however, depends on $\beta _{n}$;
 it arises from the nonvanishing coupling and the rather slow divergence of the detuning. 
These logarithmic terms cannot be factored out, unless the MS couplings $\lambda_{n}$ coincide or vanish by accident,
 and appear in some transition probabilities, as we shall see below.

\subsubsection{The propagator in the original basis}

By taking into account the LZ propagators (\ref{Un}) for the $N_{b}$ two-state MS systems, the ordering of the states,
 and the MS Hamiltonian (\ref{H-MS explicit}), the full propagator in the MS basis can be written as%
\begin{equation}
\widetilde{\mathbf{U}}=\left[ \!%
\begin{array}{cc}
\mathbf{1}_{N_{d}} & \mathbf{0} \\ 
\mathbf{0} & \!\!\!\!%
\begin{array}{cccccccc}
\alpha _{1} & 0 & \cdots & 0 & -\beta _{1}^{\ast } & 0 & \cdots & 0 \\ 
0 & \alpha _{2} & \cdots & 0 & 0 & -\beta _{2}^{\ast } & \cdots & 0 \\ 
\vdots & \vdots & \ddots & \vdots & \vdots & \vdots & \ddots & \vdots \\ 
0 & 0 & \cdots & \alpha _{N_{b}} & 0 & 0 & \cdots & -\beta _{N_{b}}^{\ast }
\\ 
\beta _{1} & 0 & \cdots & 0 & \alpha _{1}^{\ast } & 0 & \cdots & 0 \\ 
0 & \beta _{2} & \cdots & 0 & 0 & \alpha _{2}^{\ast } & \cdots & 0 \\ 
\vdots & \vdots & \ddots & \vdots & \vdots & \vdots & \ddots & \vdots \\ 
0 & 0 & \cdots & \beta _{N_{b}} & 0 & 0 & \cdots & \alpha _{N_{b}}^{\ast }%
\end{array}%
\end{array}%
\!\!\right] .  \label{U-MS}
\end{equation}%
By using the completeness relation
\begin{gather}\label{completeness}
\sum_{n=1}^{N_{b}}|\widetilde{\psi}_{n}^{a}\rangle \langle \widetilde{\psi}_{n}^{a}|+\sum_{k=1}^{N_{d}}|\widetilde{\psi}_{k}^{d}\rangle \langle 
\widetilde{\psi }_{k}^{d}|=\mathbf{1}_{N_{a}}, 
\end{gather}
it is straightforward to show that the propagator in the original basis $%
\mathbf{U}=\mathbf{S^{\dagger }}\widetilde{\mathbf{U}}\mathbf{S}$ reads 
\begin{equation}
\mathbf{U}= \left[ \begin{array}{cc}
\mathbf{1}\!+\!\sum_{n=1}^{N_{b}}(\alpha _{n}\!-\!1)|\widetilde{\psi }_{n}^{a}\rangle \langle \widetilde{\psi }_{n}^{a}| &
 -\!\!\sum_{n=1}^{N_{b}} \beta_{n}^{\ast }|\widetilde{\psi }_{n}^{a}\rangle \langle \widetilde{\psi }_{n}^{b}| \\ 
  \sum_{n=1}^{N_{b}} \beta _{n}|\widetilde{\psi }_{n}^{b}\rangle \langle \widetilde{\psi }_{n}^{a}| &
 \sum_{n=1}^{N_{b}} \alpha _{n}^{\ast }|\widetilde{\psi }_{n}^{b}\rangle \langle \widetilde{\psi }_{n}^{b}|
\end{array}\right] .  \label{U}
\end{equation}%
Note that the propagator does not depend on the decoupled states $|\widetilde{\psi }_{k}^{d}\rangle$ ($k=1,2,\ldots,N_{d}$),
 which are excluded by using Eq. (\ref{completeness}). 
This has to be expected because, owing to their degeneracy, the choice of the decoupled states is not unique:
 any superposition of them is also a zero-eigenvalue eigenstate of $\mathbf{VV}^{\dagger }$. 
Because the dynamics in the original basis must not depend on such arbitrariness,
 the propagator $\mathbf{U}$ must not depend on the decoupled states at all.

\subsubsection{Transition probabilities}

If the system starts in an arbitrary state $\vert \psi_{i}\rangle $ of the $a$ set then Eq. (\ref{U}) gives
 for the matrix elements $U_{fi}=\langle \psi_f \vert \mathbf{U}\vert \psi_i \rangle $ the expressions%
\begin{equation}
U_{fi}=\left\{ \begin{array}{cc}
 \delta _{fi}+\sum_{n=1}^{N_{b}}\left( \alpha _{n}-1\right) a_{fn}a_{in}^{\ast } & (f\in a\text{ set}), \\ 
 \sum_{n=1}^{N_{b}}\beta _{n}b_{fn}a_{in}^{\ast } & (f\in b\text{ set}),%
\end{array}\right.  \label{Ufi-a}
\end{equation}
where $a_{kn}$ and $b_{kn}$ denote the components of the MS states $%
\vert \widetilde{\psi }_{n}^{a}\rangle $ and $\vert \widetilde{\psi }_{n}^{b}\rangle $, respectively, 
\begin{subequations}\label{components}
\begin{eqnarray}
|\widetilde{\psi }_{n}^{a}\rangle &=&\left[ a_{1n},a_{2n},\ldots ,a_{N_{a}n}%
\right] ^{T},  \label{a_kn} \\
|\widetilde{\psi }_{n}^{b}\rangle &=&\left[ b_{1n},b_{2n},\ldots ,b_{N_{b}n}%
\right] ^{T}.  \label{b_kn}
\end{eqnarray}%
\end{subequations}
If the initial state $\left\vert \psi_{i}\right\rangle $ belongs to the $b$ set, we have 
\begin{equation}
U_{fi}=\left\{ \begin{array}{cc}
 -\sum_{n=1}^{N_{b}}\beta_{n}^{\ast }a_{fn}b_{in}^{\ast } & (f\in a\text{ set}), \\ 
  \sum_{n=1}^{N_{b}}\alpha_{n}^{\ast }b_{fn}b_{in}^{\ast } & (f\in b\text{ set}).%
\end{array}\right.  \label{Ufi-b}
\end{equation}

In both cases, the transition probability from state $\left\vert \psi_{i}\right\rangle $ to state $|\psi _{f}\rangle $ is 
\begin{equation}
P_{i\rightarrow f}=\left\vert U_{fi}\right\vert ^{2}.  \label{Pif}
\end{equation}

Equations (\ref{Ufi-a})-(\ref{Pif}) reveal several important features of the degenerate LZ model.

(i) The transition probability $P_{i\rightarrow f}$ is always well defined if the initial and final states belong to
 the same set of states ($a$ or $b$) because then $P_{i\rightarrow f}$ involves only the Cayley-Klein parameters
 $\alpha _{n}$, which are real and positive and hence do not have divergent phases, see Eq. (\ref{alpha_n}).

(ii) When the initial and final states belong to different sets, $\left\vert\psi _{i}\right\rangle $ to the $a$ set
 and $\left\vert \psi_{f}\right\rangle $ to the $b$ set, or vice versa, the transition probability $P_{i\rightarrow f}$ is well defined
 only if the corresponding sums in Eq. (\ref{Ufi-a}) or (\ref{Ufi-b}) reduce to one term
 (because some of the $a$ and $b$ coefficients may vanish accidentally) or if the phases of all participating $\beta _{n}$'s are the same. 
The latter may only happen accidentally if all MS couplings $\lambda _{n}$ are equal: then the phases factor out and cancel in the transition probability.

(iii) Baring accidental cases discussed in the previous point, the transition probabilities between states from different sets
 are \emph{not defined} due to the divergence of the phases of the Cayley-Klein LZ parameters $\beta _{n}$.

\subsubsection{Summary}

In summary, Eq. (\ref{U}) gives the propagator for the degenerate LZ model.
The transition probabilities can be calculated from Eqs. (\ref{Ufi-a})-(\ref{Pif}),
 which require the knowledge of the coupled MS states $|\widetilde{\psi }_{n}^{a}\rangle $
 of the $a$ set and $|\widetilde{\psi }_{n}^{b}\rangle $ of the $b$ set. 
The former are the eigenstates of $\mathbf{VV}^{\dagger }$ and the latter are the eigenstates of $\mathbf{V}^{\dagger }\mathbf{V}$. 
The knowledge of the decoupled zero-eigenvalue states $|\widetilde{\psi }_{n}^{d}\rangle $ of the $a$ set
 is not necessary for the calculation of the propagator. 
Not all transition probabilities are defined for infinite time duration because of the divergent phases of the Cayley-Klein parameters $\beta _{n}$. 
For any \emph{finite} initial and final times, though, all transition probabilities are well defined.

\subsubsection{An alternative: the Allen-Eberly-Hioe model}

In a real physical situation with degenerate levels, a more realistic
 alternative to the LZ model is the lesser known Allen-Eberly-Hioe model \cite{AE,Hioe} 
\begin{subequations}
\begin{eqnarray}
\Omega (t) &=&\Omega _{0}\text{ sech}(t/T),  \label{AE coupling} \\
\Delta (t) &=&B\tanh (t/T).  \label{AE detuning}
\end{eqnarray}%
\end{subequations}
Here the coupling $\Omega(t)$ is a bell-shaped pulse, with a characteristic width $T$. 
The detuning crosses resonance at time $t=0$ and does not diverge at infinity but tends to the finite values $\pm B$. 
The Cayley-Klein parameters for this model, including their phases, are well defined.


\section{Examples \label{Sec-example}}

\subsection{$J_{a}=2\leftrightarrow J_{b}=1$ transition}

\subsubsection{General case}

\begin{figure}[tb]
\includegraphics[width=72mm]{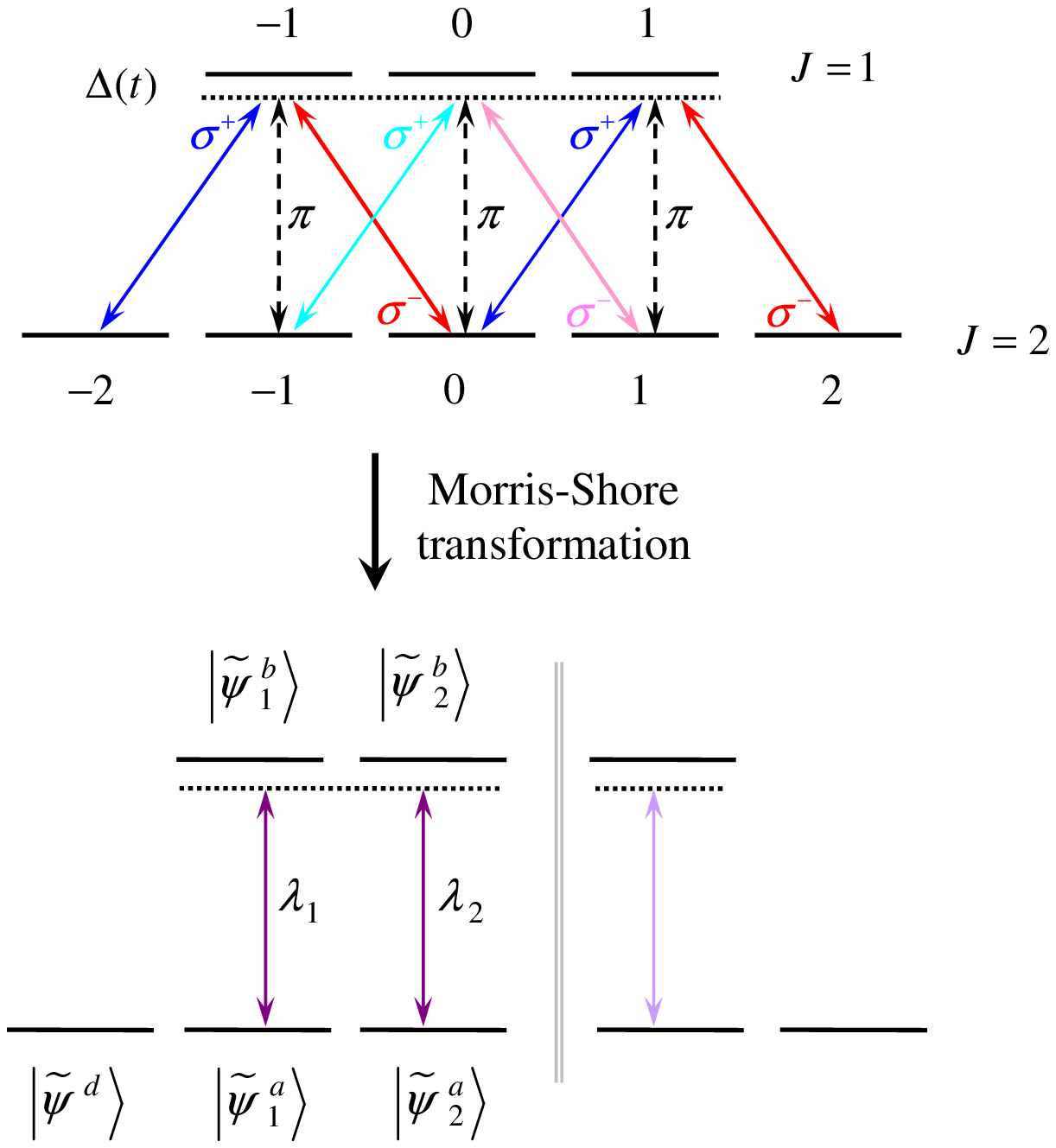}
\caption{(Color online) The $J_{a}=2\leftrightarrow J_{b}=1$ transition in
the original basis (top) and in the MS basis (bottom). With only circularly
polarized fields the full eight-state system decouples into a five-state M
system and a three-state $\Lambda $ system. A linearly polarized field would
couple the M and $\Lambda $ systems. The Morris-Shore transformation turns
the M system into a pair of two independent nondegenerate two-state systems
and a decoupled state (bottom left), and the $\Lambda $ system into a
two-state system and a decoupled state (bottom right).}
\label{Fig-M}
\end{figure}

We illustrate the above results with a specific example: transition between two atomic levels with total angular momenta
 $J_{a}=2$ and $J_{b}=1$ in the field of two circularly polarized (right $\sigma^+$ and left $\sigma^-$)
 chirped-frequency laser fields with linear chirp and steady amplitudes. 
In the absence of magnetic fields, the 5 magnetic sublevels of the $J_{a}$ level are degenerate
 and so are the 3 magnetic sublevels of the $J_{b}$ level, as shown in Fig. \ref{Fig-M}.
This system therefore represents a physical realization of the degenerate LZ model with $N_{a}=5$ and $N_{b}=3$. 
When only $\sigma ^{+}$ and $\sigma ^{-}$ polarized fields are present the eight-state system decouples into a
five-state M system, which is composed of the sublevels with $M_{a}=-2,0,2$ and $M_{b}=-1,1$,
 and a three-state $\Lambda $ system comprising the sublevels with $M_{a}=-1,1$ and $M_{b}=0$ \cite{MW}. 
If there is also a linearly ($\pi $) polarized field then the M and $\Lambda $ systems couple
 and all eight states will be involved in the dynamics.

The two $\sigma^+$ and $\sigma^-$ polarized fields can be produced by a single elliptically polarized field;
 then the amplitude ratio and the relative phase of the $\sigma^+$ and $\sigma^-$ fields can be controlled, respectively,
 by the ellipticity and the rotation angle of the field. 
Moreover the $\sigma^+$ and $\sigma^-$ fields will have automatically the same detuning.

We shall only consider the M-system, because the $\Lambda $-system contains a non-degenerate upper state
 and can be treated with a simpler formalism \cite{Kyoseva}.

\begin{table}[t]
\caption{Coefficients of the MS basis states (\protect\ref{BrightDark}) for the $J_{a}=2\leftrightarrow J_{b}=1$ transition.
The values for arbitrary elliptical polarization $\varepsilon$ are in the second column,
 and those for a linear polarization $\varepsilon=0$ in the third column.
The relevant normalization coefficient $\nu$ for elliptical polarization is listed after each group of coefficients.
}
\begin{tabular}{|c|c|c|}
\hline
& arbitrary $\varepsilon $ & $\varepsilon =0$ \\ \hline
$d^\prime_{-2}$ & $\nu _{d}\left( 1-\varepsilon \right) $ & $\sqrt{\frac{1}{8}}$ \\ 
$d^\prime_{0}$ & $-\nu _{d}\sqrt{6\left( 1-\varepsilon ^{2}\right) }$ & $-\sqrt{\frac{3}{4}}$ \\ 
$d^\prime_{2}$ & $\nu _{d}\left( 1+\varepsilon \right) $ & $\sqrt{\frac{1}{8}}$ \\ 
$\nu _{d}^{-2}$ & $4\left( 2-\varepsilon ^{2}\right) $ &  \\ \hline
$a^\prime_{-2,1}$ & $-\frac{1}{2}\nu _{1a}\left( 1+\varepsilon \right) \left(
1-6\varepsilon -\sqrt{1+24\varepsilon ^{2}}\right) $ & $\sqrt{\frac{3}{8}}$\\ 
$a^\prime_{0,1}$ & $\nu _{1a}\varepsilon \sqrt{6\left( 1-\varepsilon ^{2}\right) }$ & $\frac{1}{2}$ \\ 
$a^\prime_{2,1}$ & $\frac{1}{2}\nu _{1a}\left( 1-\varepsilon \right) \left(
1+6\varepsilon -\sqrt{1+24\varepsilon ^{2}}\right) $ & $\sqrt{\frac{3}{8}}$ \\ 
$\nu _{1a}^{-2}$ & $\sqrt{1+24\varepsilon ^{2}}\left[ \left( 1+\varepsilon
^{2}\right) \sqrt{1+24\varepsilon ^{2}}+\left( 11\varepsilon ^{2}-1\right) \right] $ &  \\ \hline
$a^\prime_{-2,2}$ & $-\frac{1}{2}\nu _{2a}\left( 1+\varepsilon \right) \left(
1-6\varepsilon +\sqrt{1+24\varepsilon ^{2}}\right) $ & $-\sqrt{\frac{1}{2}}$ \\ 
$a^\prime_{0,2}$ & $\nu _{2a}\varepsilon \sqrt{6\left( 1-\varepsilon ^{2}\right) }$ & $0$ \\ 
$a^\prime_{2,2}$ & $\frac{1}{2}\nu _{2a}\left( 1-\varepsilon \right) \left(
1+6\varepsilon +\sqrt{1+24\varepsilon ^{2}}\right) $ & $\sqrt{\frac{1}{2}}$\\ 
$\nu _{2a}^{-2}$ & $\sqrt{1+24\varepsilon ^{2}}\left[ \left( 1+\varepsilon
^{2}\right) \sqrt{1+24\varepsilon ^{2}}-\left( 11\varepsilon ^{2}-1\right) \right] $ &  \\ \hline
$b^\prime_{-1,1}$ & $\nu _{b}\sqrt{\sqrt{1+24\varepsilon ^{2}}+5\varepsilon }$ & $\sqrt{\frac{1}{2}}$ \\ 
$b^\prime_{1,1}$ & $\nu _{b}\sqrt{\sqrt{1+24\varepsilon ^{2}}-5\varepsilon }$ & $\sqrt{\frac{1}{2}}$ \\ 
$b^\prime_{-1,2}$ & $\nu _{b}\sqrt{\sqrt{1+24\varepsilon ^{2}}-5\varepsilon }$ & $\sqrt{\frac{1}{2}}$ \\ 
$b^\prime_{1,2}$ & $-\nu _{b}\sqrt{\sqrt{1+24\varepsilon ^{2}}+5\varepsilon }$ & $-\sqrt{\frac{1}{2}}$ \\ 
$\nu _{b}^{-2}$ & $2\sqrt{1+24\varepsilon ^{2}}$ &  \\ \hline
\end{tabular}
\label{Table}
\end{table}

The interaction matrix for the M system, with the Clebsch-Gordan coefficients accounted for, reads \cite{MW} 
\begin{equation}
\mathbf{V}=\frac{1}{\sqrt{10}}\left[ \begin{array}{cc}
\sqrt{6}\Omega _{+}e^{i\theta _{+}} & 0 \\ 
\Omega _{-}e^{i\theta _{-}} & \Omega _{+}e^{i\theta _{+}} \\ 
0 & \sqrt{6}\Omega _{-}e^{i\theta _{-}}%
\end{array}\right] ,
\end{equation}%
and hence the matrices $\mathbf{VV}^{\dagger }$ and $\mathbf{V}^{\dagger }%
\mathbf{V}$ are 
\begin{subequations}
\begin{equation}
\mathbf{VV}^{\dagger }=\frac{1}{10}\left[ \!\!\!%
\begin{array}{ccc}
6\Omega _{+}^{2} & \sqrt{6}\Omega _{+}\Omega _{-}e^{i\theta } & 0 \\ 
\sqrt{6}\Omega _{+}\Omega _{-}e^{-i\theta } & \Omega _{-}^{2}+\Omega _{+}^{2}
& \sqrt{6}\Omega _{+}\Omega _{-}e^{i\theta } \\ 
0 & \sqrt{6}\Omega _{+}\Omega _{-}e^{-i\theta } & 6\Omega _{-}^{2}%
\end{array}%
\!\!\!\right] ,
\end{equation}%
\begin{equation}
\mathbf{V}^{\dagger }\mathbf{V}=\frac{1}{10}\left[ 
\begin{array}{cc}
6\Omega _{+}^{2}+\Omega _{-}^{2} & \Omega _{+}\Omega _{-}e^{i\theta } \\ 
\Omega _{+}\Omega _{-}e^{-i\theta } & \Omega _{+}^{2}+6\Omega _{-}^{2}%
\end{array}%
\right] ,
\end{equation}%
with $\theta =\theta _{+}-\theta _{-}$ being the relative phase of the two
fields. The eigenvalues of $\mathbf{VV}^{\dagger }$ are $\lambda _{n}^{2}$ ($%
n=0,1,2$), where 
\end{subequations}
\begin{subequations}
\label{lambda}
\begin{eqnarray}
\lambda _{0} &=&0,  \label{L0} \\
\lambda _{1,2} &=&\Omega \sqrt{\frac{7\pm \sqrt{1+24\varepsilon ^{2}}}{20}},
\label{L12}
\end{eqnarray}%
\end{subequations}
with $\Omega =\sqrt{\Omega _{+}^{2}+\Omega _{-}^{2}}$ and $\varepsilon =(\Omega_+^2-\Omega_-^2)/(\Omega_+^2+\Omega_-^2)$. 
The eigenvalues of $\mathbf{V}^{\dagger }\mathbf{V}$ are $\lambda_1^2$ and $\lambda_2^2$. 
The eigenstates of $\mathbf{VV}^{\dagger }$ are a decoupled state $|\widetilde{\psi}^d\rangle$
 and two coupled states $|\widetilde{\psi}_1^a\rangle $ and $|\widetilde{\psi}_2^a\rangle$, which are composed of $a$ states,
 whereas the eigenstates of $\mathbf{V}^{\dagger }\mathbf{V}$ are two new $b$ states \cite{MW},
\begin{subequations}\label{BrightDark}
\begin{eqnarray}
|\widetilde{\psi }^{d}\rangle &=&\sum_{m=-2,0,2}d^\prime_{m}e^{-im\theta/2}\left\vert \psi _{m}\right\rangle ,  \label{dark}\\
|\widetilde{\psi }_{n}^{a}\rangle &=&\sum_{m=-2,0,2}a^\prime_{mn}e^{-im\theta/2}\left\vert \psi _{m}\right\rangle\quad (n=1,2),  \label{bright} \\
|\widetilde{\psi }_{n}^{b}\rangle &=&\sum_{m=-1,1}b^\prime_{mn}e^{-im\theta/2}\left\vert \psi _{m}\right\rangle \quad (n=1,2).  \label{excited}
\end{eqnarray}
\end{subequations}

The coefficients of these new MS basis states are given in Table \ref{Table} \cite{MW};
 they are related to the coefficients in Eqs. (\ref{Ufi-a})-(\ref{Pif}) as 
 $d_m = d^\prime_{m}e^{-im\theta/2}$, $a_{mn}=a^\prime_{mn}e^{-im\theta/2}$, and $b_{mn}=b^\prime_{mn}e^{-im\theta/2}$.
By using these coefficients and Eqs. (\ref{Un})-(\ref{LZ parameters}), (\ref{Ufi-a})-(\ref{Pif}), and (\ref{lambda}),
 one can find the transition probability between any two states. 

\subsubsection{Case of equal couplings}

We shall consider in some detail the special case $\Omega_{+}=\Omega _{-}$;
 then $\varepsilon =0$ and the coefficients in Table \ref{Table} simplify. 
The MS couplings (\ref{L12}), the LZ factors, and the Cayley-Klein parameters reduce to 
\begin{subequations}\label{lambdas}
\begin{eqnarray}
\lambda _{1} &=&\Omega \sqrt{\frac{4}{10}},\quad \Lambda _{1} =\frac{4}{10}\frac{\Omega ^{2}}{C}, \label{lambda1} \\
\lambda _{2} &=&\Omega \sqrt{\frac{3}{10}},\quad \Lambda _{2}=\frac{3}{10}\frac{\Omega ^{2}}{C}.  \label{lambda2}\\
\alpha _{1} &=&e^{-4\xi },\quad \beta _{1}=-e^{i\phi _{1}}\sqrt{1-e^{-8\xi }}, \\
\alpha _{2} &=&e^{-3\xi },\quad \beta _{2}=-e^{i\phi _{2}}\sqrt{1-e^{-6\xi }},
\end{eqnarray}
\end{subequations}
where $\xi =\pi \Omega ^{2}/10\beta $. 
It is particularly significant that the coefficient $a_{0,2}$ associated with state $\vert\psi_0\rangle$
 vanishes accidentally, $a_{0,2}=0$, see Table \ref{Table}. 
The implication is that the sums in Eqs. (\ref{Ufi-a}) and (\ref{Ufi-b}), which involve $a_{mn}$ coefficients,
 reduce to just single terms when state $\vert\psi_0\rangle$ is involved.
Consequently, all transition probabilities from and to state $\vert\psi_0\rangle$ are defined
 and the divergence of the phases $\phi_n$ does not show up here.

The propagator in the original basis reads (for $\theta=0$)
\begin{widetext}%
\begin{equation}
\mathbf{U}=\left[ 
\begin{array}{ccccc}
\frac18+\frac38 e^{-4\xi }+\frac12 e^{-3\xi} & -\sqrt{\frac{3}{32}}\left(1-e^{-4\xi}\right) & \frac18+\frac38 e^{-4\xi}-\frac{1}{2}%
e^{-3\xi } & -\sqrt{\frac{3}{16}}\beta _{1}^*+\frac{1}{2}\beta _{2}^* & -\sqrt{%
\frac{3}{16}}\beta _{1}^*-\frac{1}{2}\beta _{2}^* \\ 
-\sqrt{\frac{3}{32}}\left( 1-e^{-4\xi }\right)  & \frac{3}{4}+\frac{1}{4}%
e^{-4\xi } & -\sqrt{\frac{3}{32}}\left( 1-e^{-4\xi }\right)  & \sqrt{\frac{1%
}{8}\left( 1-e^{-8\xi }\right) }e^{i\varphi _{1}} & \sqrt{\frac{1}{8}\left(1-e^{-8\xi }\right) }e^{i\varphi _{1}} \\ 
\frac{1}{8}+\frac{3}{8}e^{-4\xi }-\frac{1}{2}e^{-3\xi } & -\sqrt{\frac{3}{32}%
}\left( 1-e^{-4\xi }\right)  & \frac{1}{8}+\frac{3}{8}e^{-4\xi }+\frac{1}{2} e^{-3\xi} &
 -\sqrt{\frac{3}{16}}\beta_1^*-\frac12\beta_2^* & -\sqrt{\frac{3}{16}}\beta _{1}^*+\frac{1}{2}\beta _{2}^* \\ 
 \sqrt{\frac{3}{16}}\beta _{1}-\frac{1}{2}\beta _{2} & -\sqrt{\frac{1}{8}\left( 1-e^{-8\xi }\right) }e^{-i\varphi _{1}} &
 \sqrt{\frac{3}{16}}\beta _{1}+\frac{1}{2}\beta _{2} & \frac{1}{2}%
\left( e^{-4\xi }+e^{-3\xi }\right)  & \frac{1}{2}\left( e^{-4\xi }-e^{-3\xi}\right)  \\ 
 \sqrt{\frac{3}{16}}\beta _{1}+\frac{1}{2}\beta _{2} & \sqrt{\frac18\left(1-e^{-8\xi}\right)}e^{-i\varphi _{1}} &
 \sqrt{\frac{3}{16}}\beta _{1}-\frac{1}{2}\beta _{2} & \frac{1}{2}\left(e^{-4\xi }-e^{-3\xi }\right) &
 \frac12\left( e^{-4\xi }+e^{-3\xi}\right) 
\end{array}%
\right].   \label{U=}
\end{equation}%
In the adiabatic limit $\xi \gg 1$ the matrix $\mathbf{P}=\{P_{fi}\}_{i,f=-2,0,2,-1,1}$
 with the transition probabilities $P_{i \rightarrow f}=P_{fi}$ reads
\begin{equation}
\mathbf{P}=\left[ 
\begin{array}{ccccc}
\frac{1}{64} & \frac{3}{32} & \frac{1}{64} &
 \left\vert \sqrt{\frac{3}{16}}e^{i\phi _1}-\frac{1}{2}e^{i\phi _2}\right\vert ^{2} &
 \left\vert \sqrt{\frac{3}{16}}e^{i\phi _1}+\frac{1}{2}e^{i\phi _2}\right\vert ^{2} \\ 
\frac{3}{32} & \frac{9}{16} & \frac{3}{32} & \frac{1}{8} & \frac{1}{8} \\ 
\frac{1}{64} & \frac{3}{32} & \frac{1}{64} &
 \left\vert \sqrt{\frac{3}{16}}e^{i\phi _1}+\frac{1}{2}e^{i\phi _2}\right\vert ^{2} &
 \left\vert \sqrt{\frac{3}{16}}e^{i\phi _1}-\frac{1}{2}e^{i\phi _2}\right\vert ^{2} \\ 
\left\vert \sqrt{\frac{3}{16}}e^{i\phi _1}-\frac{1}{2}e^{i\phi _2}\right\vert ^{2} & \frac{1}{8} &
 \left\vert \sqrt{\frac{3}{16}}e^{i\phi _1}+\frac{1}{2}e^{i\phi _2}\right\vert ^{2} & 0 & 0 \\ 
\left\vert \sqrt{\frac{3}{16}}e^{i\phi _1}+\frac{1}{2}e^{i\phi _2}\right\vert ^{2} & \frac{1}{8} &
 \left\vert \sqrt{\frac{3}{16}}e^{i\phi _1}-\frac{1}{2}e^{i\phi _2}\right\vert ^{2} & 0 & 0%
\end{array}%
\right] ,  \label{U=adb}
\end{equation}%
\end{widetext}

The $\beta $'s which are left over in the propagator (\ref{U=}) have divergent phases, see Eqs. (\ref{beta_n}) and (\ref{phi_n}). 
Because the respective couplings $\lambda _{1}$ and $\lambda _{2}$ [Eq. (\ref{lambdas})] are different,
 the logarithmic components in the phases of the $\beta $'s are different and therefore give rise to an interference term
 in the transition probability, which oscillates in time with a logarithmically increasing frequency.
Hence the transition probabilities with sums over different $\beta $'s do not have a limit at infinity. 
At any {\em finite} times, however, these probabilities are well defined.

\begin{figure}[tb]
\includegraphics[width=70mm]{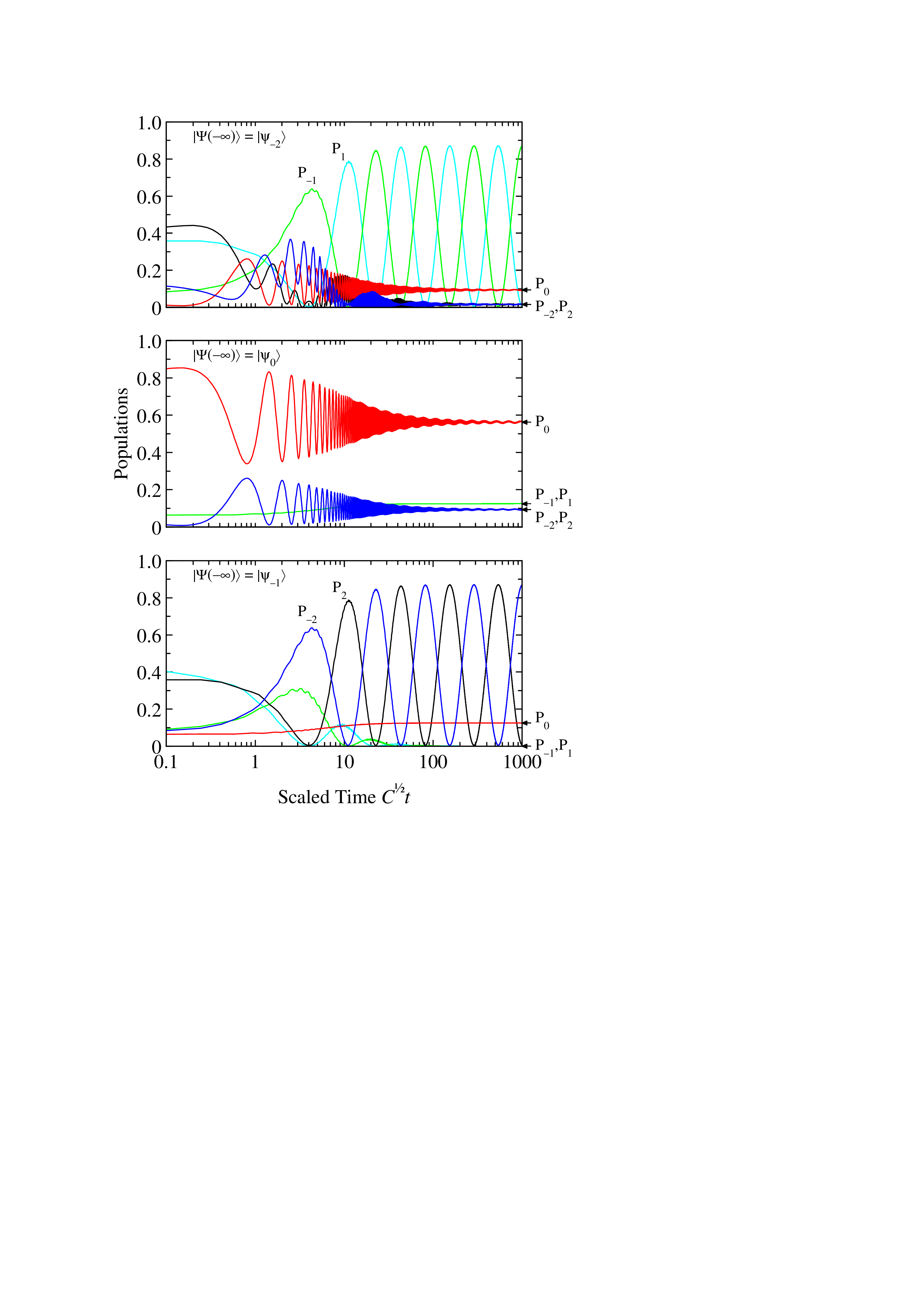}
\caption{(Color online) 
Time evolution of the populations in a five-state M system formed of the magnetic sublevels $M=-2,0,2$ of the $J=2$ level
 and $M=-1,1$ of the $J=1$ level, for linear polarization ($\varepsilon=0$), in the cases when the system starts in
 (i) top: state $\left\vert -2\right\rangle $ of the $J=2$ level; (ii) middle: state $\left\vert 0\right\rangle $ of the $J=2$ level;
 (iii) bottom: state $\left\vert -1\right\rangle $ of the $J=1$ level. 
The arrows on the right indicate the asymptotic values at $t\rightarrow \infty $, wherever applicable. 
The chirp rate $C$ is used to define the time and frequency scales. 
The coupling is $\Omega =5C^{1/2}$, which implies that the adiabatic condition ($\xi =2.5\pi \gg 1$) is fulfilled and the adiabatic solution
 (\protect\ref{U=adb}) applies. 
The initial time of the integration is $t_{i}=-400C^{-1/2}$.}
\label{Fig-populations}
\end{figure}

Figure \ref{Fig-populations} displays the time evolution of the populations of the five states in the near-adiabatic limit
 for linear polarization ($\varepsilon=0$) and for three different initial conditions. 
In the top frame the system starts in the $J=2$ state $\vert \psi_{-2}\rangle $. 
As predicted by Eq. (\ref{U=adb}) the populations of the $J=2$ states acquire definite values as $t\rightarrow \infty $,
 while the populations of the $J=1$ states oscillate: the logarithmic scale demonstrates that indeed, the oscillation phase diverges logarithmically.

Figure \ref{Fig-populations} (middle frame) displays the time evolution of the populations when the system starts in the $J=2$ state $\vert \psi_0\rangle $. 
As predicted by Eq. (\ref{U=adb}) the populations of all five states acquire definite values at infinity,
 that is all transition probabilities exist, because of the accidental vanishing of the coefficient $a_{0,2}$, as discussed above.

Figure \ref{Fig-populations} (bottom frame) displays the time evolution of the populations when the system starts
 in the $J=1$ state $\vert \psi_{-1}\rangle $. 
As predicted by Eq. (\ref{U=adb}) the populations of the $J=1$ states acquire definite values (zero) as $t\rightarrow \infty $,
 while the populations of the $J=2$ states oscillate, with a logarithmic divergence of the oscillation phase.
The exception is the population of state $\vert \psi_0\rangle $, which exists because of the accidental vanishing of the coefficient $a_{0,2}$.

\subsection{The case of arbitrary transition with $J_a=J$ and $J_b=J-1$ or $J$}

For $J_a=J$ and $J_b=J-1$ with integer $J$, in the presence of right and left circularly polarized fields only,
 the full $4J$-state system factorizes into two independent subsystems, like the M and $\Lambda $ systems in Fig. \ref{Fig-M}. 
The larger, $(2J+1)$-state system is formed of the magnetic sublevels $M_a = -J,-J+2,\ldots ,J$ of the $J_a$ level
 and $M_b=-J+1,-J+3,\ldots ,J-1$ of the $J_b$ level. 
The smaller, $\left(2J-1\right)$-state system is formed of the magnetic sublevels
 $M_a=-J+1,-J+3,\ldots ,J-1$ of the $J_a$ level and $M_b=-J+2,-J+4,\ldots ,J-2$ of the $J_b$ level.
For equally strong $\sigma^+$ and $\sigma^-$ fields ($\varepsilon=0$) the MS couplings of the larger subsystem are given in Table \ref{Table-MS}.
The smaller subsystem has the same MS couplings, except for the largest one (with $n=J$).

When $J$ is half-integer the two independent subsystems are composed of similar sets of magnetic sublevels but with opposite signs of $M$. 
Because of this symmetry, the eigenvalues are exactly the same for both subsystems.

For $J_a=J_b=J$ the two subsystems are equivalent and they have the same eigenvalues,
 which are also listed in Table \ref{Table-MS}, for both integer and half-integer $J$.

\begin{table}[t]
\caption{Morris-Shore couplings for transitions with $J_a=J$ and $J_b=J-1$ or $J$ for polarization $\varepsilon=0$.}
\begin{tabular}{|c|c|}\hline
 $J_a=J$ and $J_b=J-1$ & $J_a=J_b=J$ \\ \hline
 integer $J$ &  integer $J$ \\ 
 $\lambda_n = \Omega \sqrt{\dfrac{2n(2J-n)}{J(2J+1)}}$ & $\lambda_n = \Omega \dfrac{2n}{\sqrt{2J(J+1)}}$ \\
  ($n=0,1,\ldots ,J$) & ($n=0,1,\ldots ,J$) \\ \hline
 half-integer $J$ &  half-integer $J$ \\
 $\lambda_n = \Omega \sqrt{\dfrac{2n(2J-n)}{J(2J+1)}}$ & $\lambda_n = \Omega\dfrac{2n+1}{\sqrt{2J(J+1)}}$ \\
  ($n=0,1,\ldots ,J-1/2$) & ($n=0,1,\ldots ,J-1/2$) \\ \hline
\end{tabular}
\label{Table-MS}
\end{table}

The eigenstates (the MS states) are too cumbersome to be presented here, but they can easily be found for any particular $J$.

\section{Conclusions\label{Sec-conclusions}}

In this paper we have derived the solution of the time-dependent Schr\"{o}dinger equation
 for the degenerate Landau-Zener model, which involves two crossing sets of degenerate energies. 
The states in each set interact with the sublevels of the other set but there are no direct couplings within the same set of states. 
A physical example is the transition between the magnetic sublevels of two levels with nonzero angular momenta
 induced by steady laser fields with linearly chirped frequencies.

The solution uses the Morris-Shore transformation, which decomposes the original fully coupled system
 into a set of independent nondegenerate two-state LZ systems and a set of decoupled, dark states. 
Using the known two-state LZ propagators we use the inverse transformation to obtain the propagator in the original basis.

Our results complement the Demkov-Ostrovsky model, which assumes two crossing bands of equidistant \emph{nondegenerate} energies. 
Our results also complement the bow-tie models, which also exclude degeneracies. 
Our derivation is simpler than in these nondegenerate models;
 however, the results are not so remarkably simple, as in these models,
 because of interferences between the different LZ propagators in the MS basis.
More importantly, we have found that \emph{not all transition probabilities exist} for an \emph{infinite} coupling duration,
 because this unphysical assumption gives rise to a divergent phase in the original nondegenerate LZ model.
In the latter model the transition probability is not affected because this phase is cancelled. 
In the present degenerate LZ model, however, these divergent phases interfere and make some of the
 transition probabilities undefined in the limit of infinite times. 
As a rule, the transition probability between any two states within the same set always exists,
 but between two states from different sets can only exist by accident.

Our results demonstrate that the LZ model should be used with care when multiple states are involved.
In real physical situations the lesser known Allen-Eberly-Hioe model \cite{AE,Hioe}
 can be a viable alternative, particularly in the presence of degeneracies,
 because it involves a pulse-shaped interaction, and hence no phase divergence.


\end{document}